\documentclass[preprint]{emulateapj} 
\usepackage{courier,graphicx,epsf,color,colordvi,natbib}

\bibpunct{(}{)}{;}{a}{}{,}   

\def\kms{\hbox{km$\;$s$^{-1}$}}
\def\deg{\hbox{$^\circ$}}      
\def\Halpha{\mbox{H\hspace{0.1ex}$\alpha$}} 
\def\FeI{\ion{Fe}{1}} 
\def\CaIIH{\ion{Ca}{2}\,\,H}
\def\etal{et al.}          
\def\EB{Ellerman bomb}
\def\EBs{Ellerman bombs}

\newcommand{\avg}[1]{\left< #1 \right>} 

\def\cite#1{\citealp{#1}}    

\newcommand{\hirokoemail}{watanabe@kwasan.kyoto-u.ac.jp}

\begin{document}

\title{Ellerman Bombs at high resolution: \\  
I. Morphological evidence for photospheric reconnection}
 
\author{Hiroko Watanabe\altaffilmark{1}}
\author{Gregal Vissers\altaffilmark{2}}
\author{Reizaburo Kitai\altaffilmark{1}}
\author{Luc Rouppe van der Voort\altaffilmark{2}}
\author{Robert J. Rutten\altaffilmark{2,3}}
\affil{\altaffilmark{1}Kwasan and Hida Observatories, Kyoto
  University, Yamashina-ku, Kyoto 607-8417, Japan; \hirokoemail}
\affil{\altaffilmark{2}Institute of Theoretical Astrophysics,
  University of Oslo, %
  P.O. Box 1029 Blindern, N-0315 Oslo, Norway}
\affil{\altaffilmark{3}Sterrekundig Instituut, Utrecht University, %
  Postbus 80\,000, NL-3508 TA, Utrecht, The Netherlands}

\shorttitle{Ellerman bombs at high resolution}
\shortauthors{H. Watanabe et al.}

\begin{abstract}
  High-resolution imaging-spectroscopy movies of solar active region
  NOAA 10998 obtained with the CRisp Imaging SpectroPolarimeter
  (CRISP) at the Swedish 1-m Solar Telescope show very bright, rapidly
  flickering, flame-like features that appear intermittently in the
  wings of the Balmer \Halpha\ line in a region with moat flows and
  likely some flux emergence.  They show up at regular \Halpha\
  blue-wing bright points that outline magnetic network, but flare
  upward with much larger brightness and distinct ``jet'' morphology
  seen from aside in the limbward view of these movies.  We classify
  these features as \EBs\ and present a morphological study of their
  appearance at the unprecedented spatial, temporal, and spectral
  resolution of these observations.  
  The bombs appear along magnetic network with footpoint 
  extents up to 900\,km.  They show apparent travel away from 
  the spot along the pre-existing network at speeds of about 1\,\kms.  
  The bombs flare repetitively with much rapid variation at time
  scales of seconds only, in the form of upward jet-shaped brightness
  features.  These reach heights of 600--1200\,km and tend to show
  blueshifts; some show bi-directional Doppler signature, and some
  seem accompanied with an \Halpha\ surge.  They are not seen in the
  core of \Halpha\ due to shielding by overlying chromospheric
  fibrils.  The network where they originate has normal properties.
  The morphology of these jets strongly supports deep-seated
  photospheric reconnection of emergent or moat-driven magnetic flux
  with pre-existing strong vertical network fields as the mechanism
  underlying the \EB\ phenomenon.
 \end{abstract}


\section{Introduction}\label{sec:introduction}

\EBs\ are prominent, sudden, short-lived brightness
enhancements of the outer wings of strong optical lines, in particular
the Balmer \Halpha\ line at $\lambda = 6563$\,\AA, that occur in solar
active regions.  In his discovery paper,
\citet{1917ApJ....46..298E} 
described them as ``a very brilliant and very narrow band extending
four or five angstroms on either side of the [\Halpha] line, but not
crossing it'', fading after a few minutes.  He named them ``solar
hydrogen bombs''; nowadays, they are called after him.  They occur
exclusively in emerging flux regions, are very bright, and have a
characteristic elongated shape at sufficient angular resolution.

The fairly extensive literature on \EBs\ is reviewed in a parallel
paper (Rutten \etal, in preparation; henceforth Paper~I) which also
addresses the issue that not all bright features in the wings of
\Halpha\ are necessarily \EBs\ (or ``moustaches'', often taken as
another name for the same phenomenon).  Small photospheric
concentrations of strong magnetic field of the type that constitute
network and plage can also produce strikingly bright points in the
wings of \Halpha, the blue wing in particular 
\citep{2006A&A...449.1209L, 2006A&A...452L..15L}.
We refer to Paper~I for this issue and for a wider overview of the
\EB\ literature.  The older part was summarized well in the
introduction of
\citet{2002ApJ...575..506G}. 
A recent review of small-scale photospheric magnetic fields is given
by \citet{2009SSRv..144..275D}. 

Selected recent Ellerman-bomb studies of particular relevance here, 
in addition to the comprehensive study of 
\citet{2002ApJ...575..506G}, 
are those by Pariat \etal\
\citep{2004ApJ...614.1099P, 2007A&A...473..279P}, 
\citet{2007ApJ...657L..53I}, 
\citet{2008ApJ...684..736W}, 
\citet{2008PASJ...60...95M}, 
\citet{2010PASJ...62..879H}, 
\citet{2010PASJ...62..901M}, 
and
\citet{2010ApJ...724.1083G}. 
These studies provide evidence that the \EB\ phenomenon occurs at
sites of and are due to magnetic reconnection, and characteristically
have elongated shapes reminiscent of the ``chromospheric anemone
jets'' of
\citet{2007Sci...318.1591S}. 
The present paper strengthens this case on the basis of \Halpha\
observations that are far superior in quality to any in the
literature, and establishes that \EBs\ of the type seen here are
rooted in the deep photosphere.  In the remainder of this introduction
we show and discuss a few illustrative examples, and then report more
comprehensive measurements in the main body of the paper.

\begin{figure*}[bhtp]
  \centerline{\includegraphics[width=\textwidth]{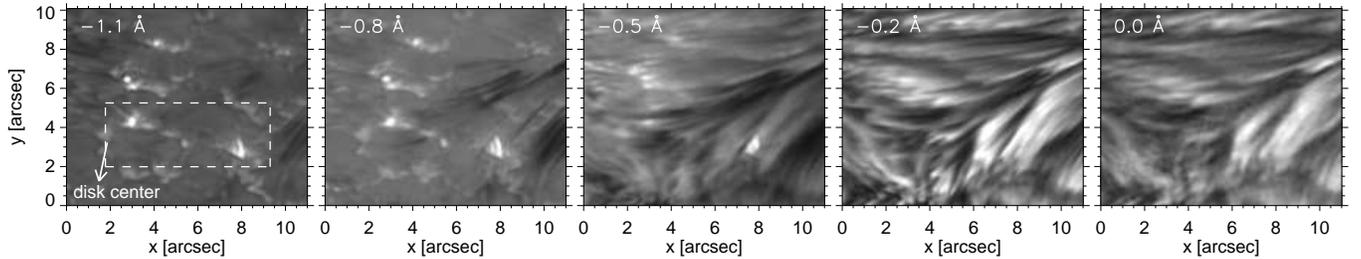}}
  \caption[]{\label{fig:cut-wavs}%
    A sample of \EBs\ in the small subfield of the CRISP profile scan
    taken at 08:01\,UT 
    that is outlined in the upper-left panel of
    Figure~\ref{fig:full-fov}.  Clockwise rotation over 120\deg\ from
    $(X,Y)$ was applied to obtain local $(x,y)$ coordinates with the
    limbward (vertical) direction nearly upward; the direction to disk
    center is shown in the first panel.  The spot lies to the left;
    nearby plage lies to the right.  The five images sample \Halpha\
    at $\Delta \lambda = -1.1, -0.8, -0.5, -0.2, 0.0$\,\AA\ from line
    center.  Each panel is greyscaled separately.  
    The dashed white frame in the first panel outlines the yet smaller 
    subfield used in Figure~\ref{fig:cut-time}.}
\end{figure*}

\begin{figure*}[bhtp]
\centerline{\includegraphics[width=0.9\textwidth]{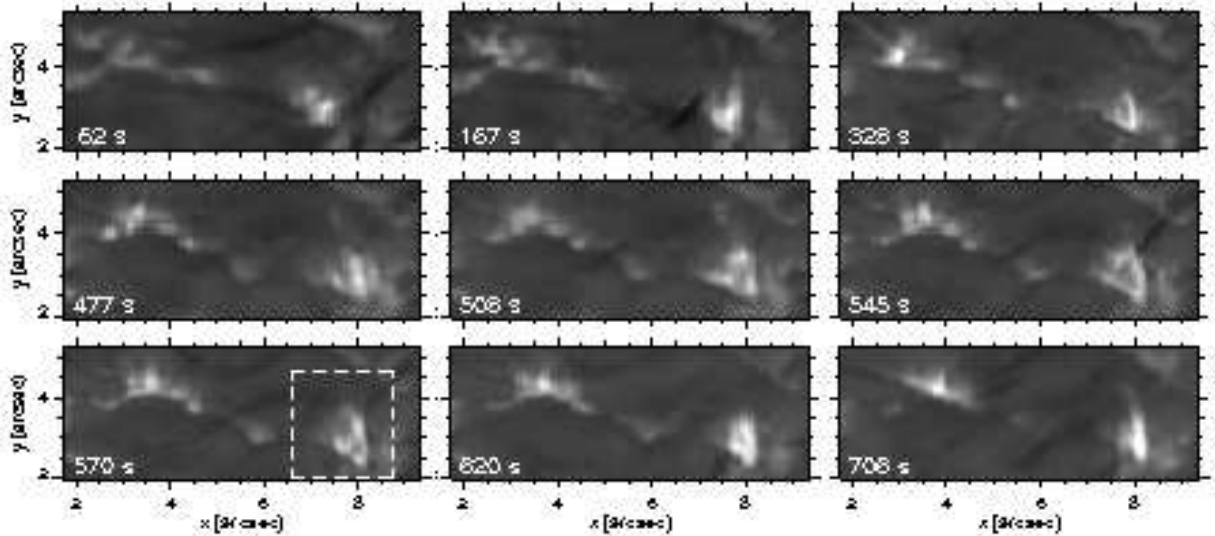}}
\caption[]{\label{fig:cut-time}
  \EB\ evolution in a selection of CRISP image cutouts at $\Delta
  \lambda = -1.1$\,\AA\ from \Halpha\ line center.  The small subfield
  shown here is outlined in Figure~\ref{fig:cut-wavs} and contains
  the most prominent \EB\ (EB-1) at $(x,y)=(8,3)$.  
  The elapsed time from 07:55\,UT is specified in each panel and
  increases in unequal steps along rows.  The $t=328$\,s panel
  corresponds to the samples in Figures~\ref{fig:cut-wavs},
  \ref{fig:full-fov}, and \ref{fig:ha-profiles}.  The dashed white
  frame specifies the EB-1 cutout shown at 2-s cadence in
  Figure~\ref{fig:cut-fast}.  High resolution figure is available at 
  http://www.kwasan.kyoto-u.ac.jp/\~{}watanabe/onlinematerial/f2.eps
}
\end{figure*}

\begin{figure*}[bhtp]
  \centerline{\includegraphics[width=\textwidth]{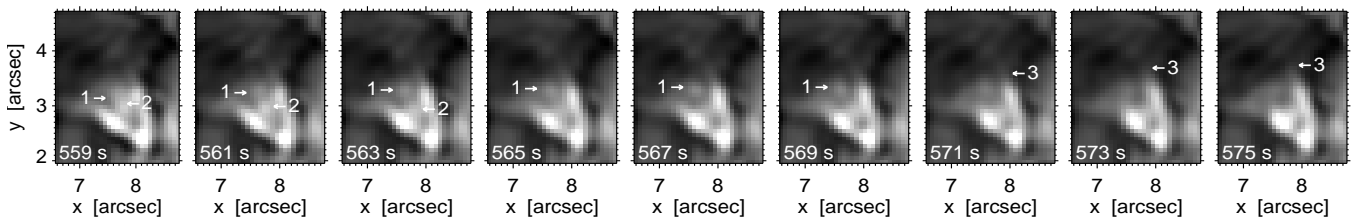}}  
  \caption[]{\label{fig:cut-fast}%
    Rapid variations in \EB\ EB-1 in simultaneously taken wide-band
    images at 2-s cadence.  The subfield contains only EB-1.  It is
    outlined in the 7th panel of Figure~\ref{fig:cut-time} which is
    straddled by the brief sequence shown here.  The elapsed time
    specified in the lower-left corner of each panel corresponds to
    that in Figure~\ref{fig:cut-time}. Upward motion
    of brightness features is marked co-moving arrow 1;						    
    downward motion with arrow~2, and extension of an elongated
    jet structure with arrow~3.  }
\end{figure*}

Our observations, described in detail in
Section~\ref{sec:observations}, were obtained with the CRisp Imaging
SpectroPolarimeter (CRISP) at the Swedish Solar 1-m Telescope (SST) on
La Palma.  The target was an active region with a spot and extended
plage where flux was still emerging, seen limbward at viewing angle
$\mu =0.67$. A 37-minute sequence of high-resolution images sampling
23 wavelengths covering \Halpha\ was taken at a temporal cadence of
6.2~s between consecutive \Halpha\ profile scans.

The observations can be inspected as movies in the online material 
(available at http://www.kwasan.kyoto-u.ac.jp/\~{}watanabe/onlinematerial/).
They show a few dozen eruptive features that are very bright in the
\Halpha\ wings and appear as rapidly varying extended flames or jets,
jutting out from photospheric network, pointing away from disk center. 
These are clearly \EBs, seen at unprecedented spatial and temporal
resolution.  The limbward viewing helps to display their upward extent
and fast variations as seen from the side.  It also diminishes the
network bright point contrast so that the bombs stand out more clearly
(cf.\,Paper~I).  Figures~\ref{fig:cut-wavs}--\ref{fig:cut-fast}
discussed below show cut-out snapshots that serve as a quick-look
emulation of viewing the \EBs\ in these movies, but we strongly
recommend that the reader inspects the movies.

\paragraph{Ellerman bomb visibilty}
Figure~\ref{fig:cut-wavs} demonstrates, for a small subfield, why
\EBs\ appear only in the outer wings of \Halpha\ which sample the deep
photosphere, apart from rare Doppler-shifted fibrils
(\cite{2006A&A...449.1209L}; 
Paper~I).  Closer to line center, they are fully shielded from view by
the overlying opaque \Halpha\ chromosphere made up of dark fibrils.
Note that each panel is greyscaled separately for best contrast; in
intensity units, the brightest features at line center are much darker
than the \EBs\ in the wings, so these do not poke through the fibril
canopy even though they extend to heights normally considered
chromospheric.  Comparing the line-wing and line-center movies in the
online material confirms that the line-center fibril evolution is
unrelated to the \EB\ activity underneath, excepting the few \EBs\
that are accompanied by an \Halpha\ surge.

\paragraph{Ellerman bomb evolution}
Figure~\ref{fig:cut-time} samples the time evolution of the yet
smaller subfield specified in Figure~\ref{fig:cut-wavs}.  The largest
\EB\ is designated EB-1 henceforth.  It migrates to the right, along
the network away from the spot, with consecutive flaring-up of
adjacent threads that reach large extent with apparently near-vertical
orientation.  At some instances, for example at $t=328$\,s, the bright
threads have thicker bright feet suggesting small-scale ``inverted Y''
(or ``Tour d'Eiffel'') morphology.  The \EB\ site at $(x,y)=(3.5, 4.5)$ shows a
similar sequence of smaller thread-like brightenings migrating to the
right along the network.

Figure~\ref{fig:cut-fast} samples even faster variations of EB-1 in
frames taken simultaneously with the CRISP sequence but at much higher
cadence, with another camera registering \Halpha\ with a 4.9\,\AA\
wide passband. 
In order to enhance the contrast of interesting features in the still images, 	
an unsharp masking method similar to the procedure described by 
\citet{2010PASJ...62..879H}, 
but using a Gaussian filter width of 0.5~arcsec, was employed.
The figure demonstrates that small-scale brightness features
within EB-1 change and move on timescales of seconds.  The apparent
speeds are measured in Section~\ref{sec:results} and are supersonic.

\paragraph{Ellerman bomb properties}
Viewing the movies establishes a number of key
properties of the \EBs\ in these data.  They are:
\begin{itemize}  

\item 
These \EBs\ are all rooted in intergranular lanes in the deep
photosphere;

\item They all have upward protrusions that we generically call
  ``jets'' here, often appearing in the form of nearly
  parallel or consecutive slender bright threads;

\item 
  They all point away from disk center in the limbward field of view,
  and so appear to be more or less vertically oriented.  None seems to
  sense the more horizontal orientation of the overlying chromospheric
  fibrils that are observed at the center of \Halpha;

\item The \EBs\ migrate, while flaring repetitively, along the
  pre-existing network, always away from the spot;

\item There are very fast variations in their fine structure during
  the repetitive flaring;

\item Some of the brighter threads  display small-scale
   inverted-Y morphology;

\item Some \EBs\ are accompanied by dark \Halpha\ surges. 

\end{itemize}

Together, these properties define the \EBs\ in this data set
as photospheric events rooted in the network that produce eruptive
jets reaching upward to considerable height, with fast temporal
variations that suggest successive interaction of newly arriving
vertical field with existing network field, most likely in the form of
magnetic reconnection.  Photospheric reconnection has been suggested
theoretically (cf.\,
\cite{1982SoPh...77..121S}; 
\cite{1995Natur.375...42Y}; 
\cite{1999ApJ...515..435L}) 
but has not been estabished observationally.  The present observations
provide the clearest evidence to date.

In the rest of the paper we detail the observations
(Section~\ref{sec:observations}) and the methods used for the
selection and measurements of \EBs\ and other bright points
(Section~\ref{sec:measurements}).  In Section~\ref{sec:results} we
quantify the observed Ellerman-bomb footpoint motions, extensions and
retractions, their Doppler signatures, and their fast variations.  We
also compare Ellerman-bomb properties with those of non-eruptive
network bright points and display one of the two cases with a surge.
We discuss the \EB\ phenomenon in Section~\ref{sec:discussion},
including a brief comparison with previous studies, and conclude the
paper in Section~\ref{sec:conclusion}.

\begin{figure*}[bhtp]
  \centerline{\includegraphics[width=\textwidth]{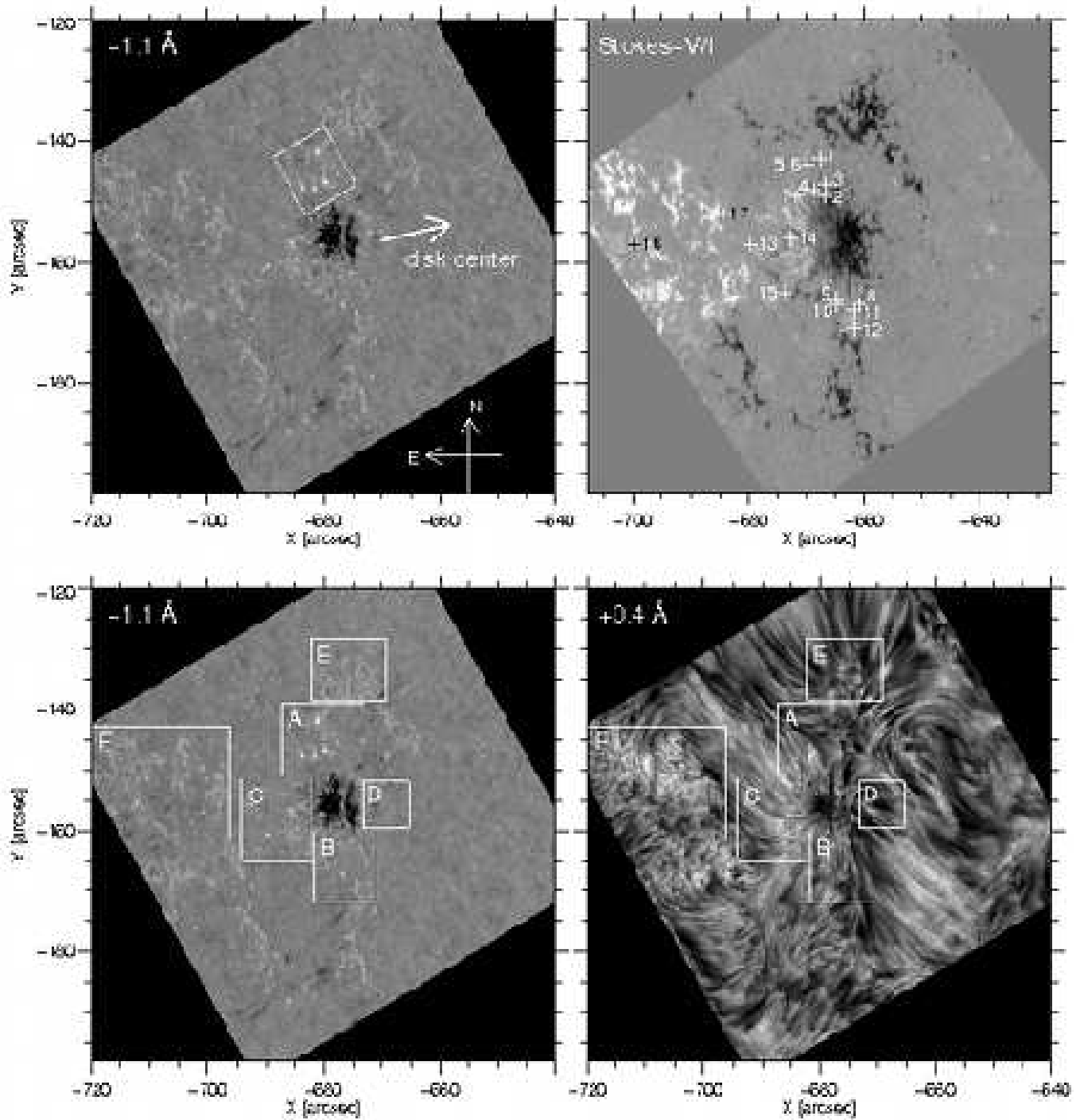}}
  \caption[]{\label{fig:full-fov}%
    CRISP image samples for the full field of view. The standard solar
    $(X,Y)$ coordinates have the origin at apparent disk center and
    the Y direction towards the solar poles with northward positive
    and the X direction with westward positive.  The center of the
    field of view has viewing angle $\mu=0.67$ towards the east limb;
    the direction towards disk center is indicated in the first
    panel. The field of view includes AR\,10998.  {\it Upper-left
      panel\/}: \Halpha$-1.1$\,\AA\ blue-wing image taken at 08:01~UT.
    The white frame specifies the subfield that is shown in
    Figure~\ref{fig:cut-wavs} with clockwise rotation over 120\deg.
    The brightest feature in this subfield is \EB\ EB-1, shown in
    detail in Figures~\ref{fig:cut-time} and \ref{fig:cut-fast}.  {\it
      Upper-right panel\/}: a CRISP \FeI~6302\,\AA\ Stokes-$V/I$
    magnetogram taken two hours later, with corresponding shift in $X$
    due to solar rotation.  The numbered plus signs define the locations
    of the 17 \EBs\ measured in Section~\ref{sec:measurements}.  {\it
      Lower-left panel\/}: the same \Halpha$-1.1$\,\AA\ blue-wing
    image overlaid with frames specifying the subregions selected for
    bright-point measurements in Section~\ref{sec:measurements}.  {\it
      Lower-right panel\/}: simultaneous \Halpha$+0.4$\,\AA\ image
    overlaid with the same subregion frames. High resolution figure is available at 
  http://www.kwasan.kyoto-u.ac.jp/\~{}watanabe/onlinematerial/f4.eps}
\end{figure*}

\section{Observations and Data Reduction}\label{sec:observations}

\paragraph{Instrumentation}
The data for this study were obtained with the CRisp Imaging
SpectroPolarimeter 
(CRISP; \cite{2008ApJ...689L..69S}) 
at the Swedish 1-m Solar Telescope
(SST; \cite{2003SPIE.4853..341S}) 
on La Palma.  CRISP is a dual Fabry-P\'erot interferometer allowing
wavelength tuning within 50\,ms.  Its combination with the
unprecedented angular resolution and image sequence quality attained
by the SST's superb optics and site combined with sophisticated
wavefront restoration yields data of outstanding spatial, temporal and
spectral resolution (e.g., 
\cite{2009ApJ...705..272R}; 
\cite{2010ApJ...713.1282O}). 
The SST real-time tip-tilt correction and adaptive-optics wavefront
correction are described by
\citet{2003SPIE.4853..370S}. 
The post-detection numerical image restoration is described by
\citet{2005SoPh..228..191V}. 

\paragraph{Observations}
The setup of CRISP starts with an optical chopper synchronizing the
exposures by multiple cameras. A filterwheel contains appropriate
prefilters; in this case one was used that selects \Halpha\ with a
passband of ${\rm FWHM} = 4.9$\,\AA.  A few percent of the light is
branched off to a camera taking images in this wide band.  These serve
usually only for ``multi-object'' seeing monitoring in the numerical
post-processing, but in this study they were also analyzed because of
their fast cadence.  The bulk of the light passes through two liquid
crystals and a high-resolution and a low-resolution Fabry-P\'erot
etalon, and is then divided by a polarizing beam splitter between two
cameras.  The three CCD cameras are high-speed low-noise Sarnov
CAM1M100 ones with 1K$\times$1K chips.  They ran synchronously at 35
frames per second with exposure time 17\,ms.  The image scale was
0.071~arcsec\,px$^{-1}$, well sampling the SST's Rayleigh diffraction
limit of 0.17~arcsec at \Halpha.  The field of view covered
$67\times67$\,arcsec$^2$.  The CRISP transmission profile has ${\rm
  FWHM} = 6.6$\,pm at \Halpha.

An \Halpha\ sequence was acquired between 07:55 and 08:33\,UT on 2008
June 11 during good to excellent seeing conditions.  The target was
AR\,10998, containing a leading sunspot with a rudimentary penumbra
and following plages.  The sunspot appeared on June 8 2008, kept
growing until June 11, and had fully disintegrated 
by June 15.  In these June 11 data, the region was located at viewing angle $\mu=0.67$
towards the East limb.  The field of view was centered on $(X,Y)
\approx (-684, -152)$ with $(X,Y)$ the standard solar coordinates in
arcsec from apparent disk center with $Y$ positive along the solar
meridian towards the North pole, $X$ positive westward.

The \Halpha\ profile was sequentially sampled at 23 wavelengths within
the line, ranging between $\pm 1.1$\,\AA\ with respect to line center
with 0.1\,\AA\ spacing.  At each tuning position a ``multi-frame''
burst of 8 exposures was taken.  The cadence between consecutive line
scans was 6.2\,s.

Another observing program was run before and after the \Halpha\
observations, targeting the same area but sampling the magnetically
sensitive \FeI~6302\,\AA\ lines.  It ran during 07:15--07:32\,UT and
09:26--10:26\,UT.  The first run had CRISP sampling both the
6301.5\,\AA\ and 6302.5\,\AA\ lines as well as a continuum wavelength,
at a scan cadence of 41\,s.  The later run sampled only the
6302.5\,\AA\ line plus a continuum wavelength at a scan cadence of
16\,s.  Simple Stokes $V/I$ magnetograms were constructed from the
flat-fielded images at $\Delta \lambda = -48$\,m\AA\ from line center,
selected for high Stokes-$V$ signal and large contrast, for inspection
of the magnetic context prior to and after the \Halpha\ observations.
One such magnetogram is shown in Figure~\ref{fig:full-fov}.

\paragraph{Reduction}
The post-processing to achieve further image restoration including
precise co-alignment was done with the Multi-Object Multi-Frame Blind
Deconvolution (MOMFBD) algorithm of
\citet{2005SoPh..228..191V}. 
It includes tesselation of all images (i.e., at each tuning position
within a line scan) into $64\times64$\,px$^2$ subfields for individual
restoration.  The wide-band data served both as multi-object
representation and as anchor for alignment of the narrow-band bursts
per sequence.  In this case the wide-band frames were also processed
separately from the CRISP data to obtain a simultaneous sequence at a
cadence of only 1\,s, which serves below to analyze fast temporal
evolution within \EB\ EB-1 (Figure~\ref{fig:cut-fast}).  More detail
on MOMFBD reconstruction strategies for such data sets is given in
\citet{2008A&A...489..429V}. 

Subsequently, the restored images were subjected to prefilter
correction removing the transmission profile of the wide-band filter,
to derotation which removes the time-varying image rotation that
results from the alt-azimuth configuration of the SST, and to
destretching which removes remaining rubber-sheet distortions
(small-scale translations and warping) by seeing.  These are
determined following
\citet{1994ApJ...430..413S} 
through cross-correlation of running means over a few minutes of small
subfields of the wide-band sequence, and are then applied to the
narrow-band images.

Figure~\ref{fig:full-fov} shows two full-field samples, for the outer
blue wing at $\Delta \lambda = -1.1$\,\AA\ from line center in the
lefthand column and for the inner red wing at $\Delta \lambda =
+0.4$\,\AA\ from line center in the lower right panel, respectively.
Both are from the CRISP scan at 08:01~UT, a moment of very good seeing
that yielded the largest image contrast of the 37-minute sequence.
The upper right panel contains a corresponding magnetogram from the
\FeI~6302\,\AA\ data taken two hours later.  
The various figure annotations are described below. 

\section{Feature selection and measurement} \label{sec:measurements}

\subsection{Ellerman bombs} \label{sec:EBs}

\paragraph{Occurrence in the field of view}
Inspection of the various \Halpha-wing movies shows the occurrence of
many \EBs\ near the spot, in particular to the East of it where there
was an extended plage of opposite polarity (upper panels of
Figure~\ref{fig:full-fov}).  Note that the penumbra shows
black-and-white bipolar difference between the East and West side of
the spot, but this results from the combination of slanted viewing and
near-horizontal penumbral field configuration.  The plage fields are
primarily vertical; their black/white signature in the magnetogram in
Figure~\ref{fig:full-fov} indeed specifies opposite polarity.  Most
\EBs\ appear fairly close to the spot.

\begin{figure}[bhtp]
  \centerline{\includegraphics[width=0.9\columnwidth]{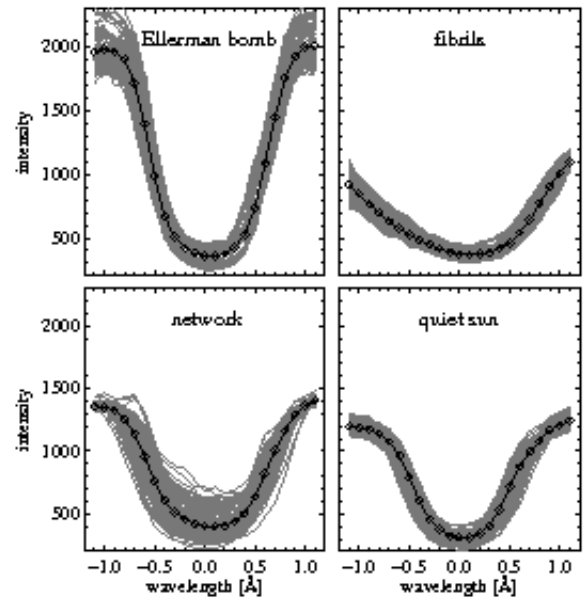}}
  \caption[]{\label{fig:ha-profiles}%
    Characteristic \Halpha\ profiles collected from the CRISP scan
    taken at 08:01\,UT.  Each panel contains a number of per-pixel
    profiles for a specific pixel category.  The solid curve is their
    mean, with the CRISP sampling wavelengths marked.  The intensity
    scale (in arbitrary data units) is the same for the different
    categories.  The samples were selected by first obtaining the
    average wing brightness $\avg{I_{\rm wing}}$ per pixel by summing
    the blue- and red-wing values at $\Delta \lambda = -1.1, -1.0,
    -0.9, +0.9, +1.0, +1.1$\,\AA.  {\it Clockwise from top left\/}: 
    \EBs, all 84 pixels with $\avg{I_{\rm wing}} > 1800$; fibrils, 314
    pixels with $\avg{I_{\rm wing}} < 950$ in the Eastern part of the 
    field of view; quiet Sun, 250 pixels covering a small area in the
    region around $(X,Y)=(-665,-170)$ in
    Figure~\ref{fig:full-fov}; network without \EBs, all 169 pixels
    with $1349 < \avg{I_{\rm wing}} < 1350$.  Note that at the $\Delta
    \lambda = \pm 1.1$\,\AA\ sampling extremes the mean profile reaches
    only about 70\% of the continuum intensity for a quiet-Sun average. 
    High resolution figure is available at 
  http://www.kwasan.kyoto-u.ac.jp/\~{}watanabe/onlinematerial/f5.eps}
\end{figure}

\paragraph{Spectral variations}
Figures~\ref{fig:cut-wavs} and \ref{fig:ha-profiles} illustrate the
spectral behavior of \Halpha\ over the target area.
Figure~\ref{fig:cut-wavs} samples the small subfield outlined in Figure~\ref{fig:full-fov}                                                               
at five wavelengths in the blue \Halpha\ wing.  The wing images ($\Delta
\lambda=-1.1$\AA, $-0.8$\AA) show a grayish background with somewhat
brighter network.  The photospheric contribution to this part of
\Halpha\ originates in the deep photosphere, but in the blue wing the
granular contrast is flattened as explained by
\citet{2006A&A...449.1209L}. 
The network contrast diminishes towards the limb, as explained in
Paper~I.
The \EBs\ appear very bright.  Figure~\ref{fig:ha-profiles}, which
samples \Halpha\ profiles for different pixel categories, shows that
in \EBs\ the \Halpha\ wings reach nearly twice the intensity they
have in quiet-sun areas, and extend well beyond our $\Delta \lambda =
\pm 1.1$\,\AA\ sampling range -- in accordance with Ellerman's
\citep{1917ApJ....46..298E} 
remark that they extend 4--5\,\AA\ on either side.

At line center all \EBs\ are hidden by overlapping chromospheric
fibrils (or ``mottles'', we use the term fibril generically for long
slender filamentary structures).  The second panel of
Figure~\ref{fig:cut-wavs} shows the onset of this obscuration.  At
$\Delta \lambda = -0.8$\,\AA\ only a few dark fibrils appear, as was
also the case in the observations of
\citet{2006A&A...449.1209L} 
at the same wavelength.  These isolated dark fibrils are mostly
chromospheric absorption features with large blueshift, probably
``Rapid Blue Excursions'' that
\citet{2009ApJ...705..272R} 
identified as the on-disk counterpart of so-called spicules-II seen at
the solar limb.  Closer to line center, the chromospheric fibrils gain
opacity and together form an opaque blanket obscuring any photospheric
contribution underneath, including the upright jets of the \EBs.  At
$\Delta \lambda = -0.5$\,\AA\ all brightest areas in
Figure~\ref{fig:cut-wavs} still correspond to the underlying \EBs,
although not 1:1 in shape, but at $\Delta \lambda = -0.2$\,\AA\ this
is no longer true.  Even the tallest \EB\ in
Figure~\ref{fig:cut-wavs}, EB-1, is obscured at line center.  

Correspondingly, the \Halpha\ cores in Figure~\ref{fig:ha-profiles}
are as dark for the \EB\ locations as they are for dark fibrils that
are selected as such on the basis of dark outer wings. Obviously, the
fibril obscuration determines this part of the \EB\ profiles.  Since
\EBs\ always occur in emerging flux regions that necessarily possess
rich fibril structure in the overlying \Halpha\ chromosphere,
obscuration-free \EB\ observation at \Halpha\ line center may 
inherently be impossible.

\paragraph{Selection}
Our movie inspections suggest that the \EBs\ are relatively long-lived
brightness entities, appearing repetitively while their feet migrate
along the network.   
We decided to use the outer \Halpha\ wings as
measurement diagnostic by defining their sum and difference as $
I_{\rm sum} \equiv (I_{r}+I_{b})/2 $ and $ I_{\rm diff} \equiv (I_{r}
- I_{b})/(I_{r}+I_{b}) $, with $I_{b}$ the spectrally averaged
blue-wing intensity at $\Delta \lambda = -1.1, -1.0$ and $-0.9$\,\AA\
from line center and likewise, $I_{r}$ the averaged red-wing intensity
at $\Delta \lambda = +1.1, +1.0$ and $+0.9$\,\AA\ from line center.
The outer-wing averaging reduces noise from seeing variations.  The
wing summing and differencing removes, to first order, the effect of
Dopplershift on $I_{\rm sum}$ and the effect of intensity change on
$I_{\rm diff}$.  We treat the wing difference as a Doppler signal below.       
This is a better choice for \EBs\ than measuring the profile-minimum
shift or core bisector shift which are set by the overlying fibrils.
For emission features such as \EBs,         
positive $I_{\rm diff}$ implies
redshift.  Since the \EBs\ appear to extend upright, the measured
redshifts imply downflows at the $\mu=0.67$ viewing angle.

We constructed 
$I_{\rm sum}$ wing-brightness and $I_{\rm diff}$
wing-Dopplergram images for the whole data sequence.  We used the
$I_{\rm sum}$ sequence to identify and select \EBs\ manually, using excess 
wing brightness, presence of a jet-like protrusion, and 
visibility above 240\,s duration as criteria.  Thus we selected 17 \EBs\
whose positions    
are indicated in the second panel of
Figure~\ref{fig:full-fov}.  The first fifteen lie in the periphery of
the spot or the moat around it (recognizable from the outward moat
flow across it in the movies).  EB-16 and EB-17 lie at the edge of a 
plage.  Overall, these \EBs\ seem to be located preferentially at the
network and moat boundaries.  We found no \EBs\ at the more quiescent
West side of the spot.

\begin{figure*}[htbp]
  \centerline{\includegraphics[width=0.95\textwidth]{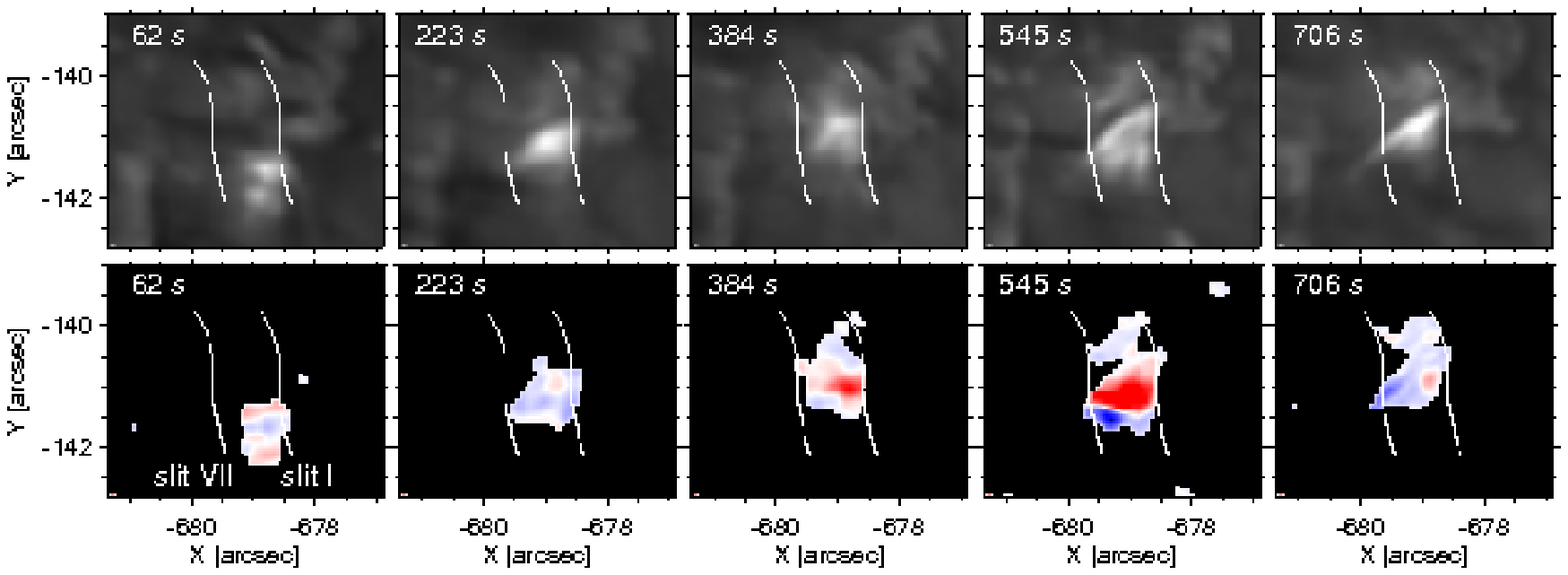}}
  \centerline{\includegraphics[width=0.95\textwidth]{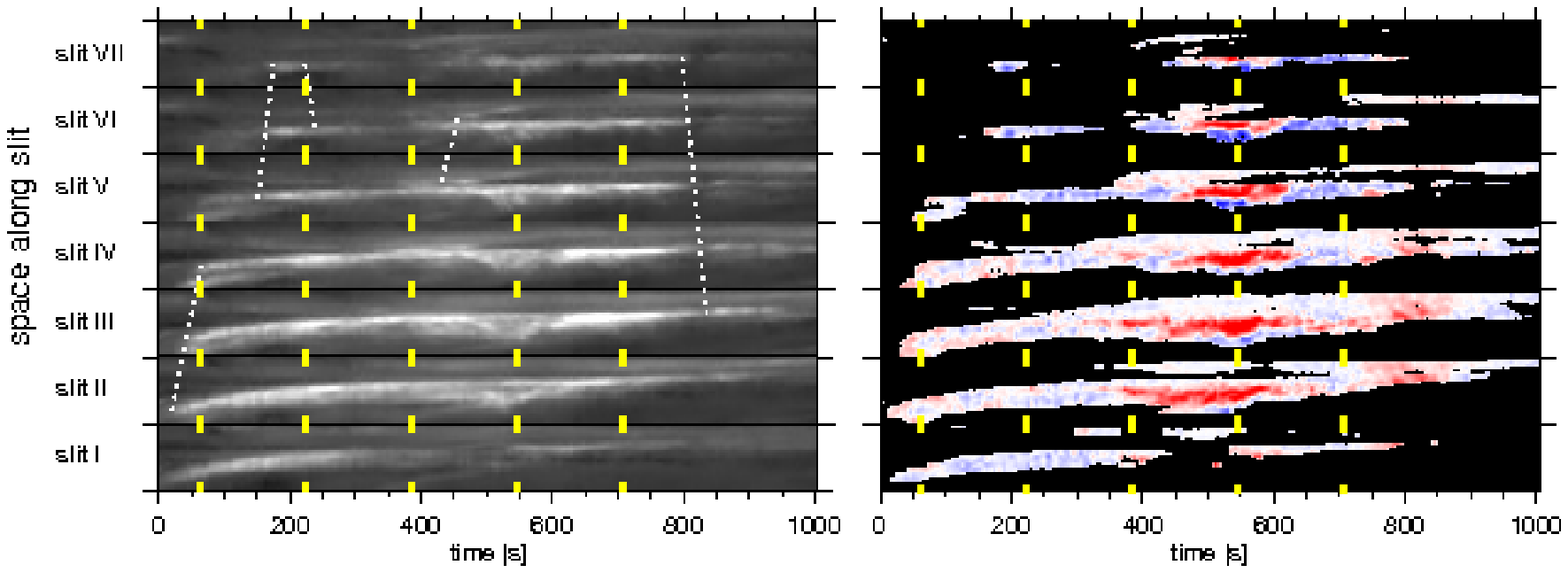}}
  \caption[]{\label{fig:slit-image}%
    {\it Upper part\/}: Wing-brightness sum ({\it upper row\/}) and difference
    (``Dopplergrams'', {\it lower row\/}) images for \EB\ EB-1.  The wing
    Dopplergram images are blue for apparent blueshift, red for
    redshift, and are thresholded at high wing-brightness intensity to
    avoid contamination with the Dopplershift signals of overlying
    dark fibrils.  The numbers in each panel specify the elapsed time
    from the first appearance at 07:55\,UT, as in
    Figure~\ref{fig:cut-time}.  The $(X,Y)$ coordinates correspond to
    Figure~\ref{fig:full-fov}; the limb is leftward.  The two white
    curves specify the extremes of seven equidistant adjacent sampling
    ``slits'' with width 1~pix.  Slit~I is set manually to track the
    footpoint location with time; the others are set parallel to that
    at intervals of 128~km in the limbward (vertical) direction.  {\it
      Lower part\/}: corresponding multi-slit space-time evolution
    diagrams for EB-1, with the elapsed time from the first appearance
    plotted horizontally and the length along each sampling slit
    plotted vertically, respectively for the wing-brightness sum
    ({\it left}\/) and difference ({\it right\/}). 
     The yellow ticks are indicators of times of snapshots 
     shown in the upper two rows, i.e., 62\,s, 223\,s, 384\,s, 545\,s, and 706\,s. 
     The dotted white lines in the
    first diagram illustrate ascents and descents of jet brightness
    features.  For example, the longest dotted line at $t=$800--840\,s
    describes contraction of the jet measured as descent of a bright
    feature at its top. The Dopplershift coding in the righthand
    diagram indicates a bi-directional jet with upward motion
    (blueshift) of its upper part and downward motion (redshift) of
    its lower part during $t=$500--750\,s. Movie versions of this
    figure and the corresponding ones for the other 16 measured \EBs\
    are available in the online material. High resolution figure is available at 
  http://www.kwasan.kyoto-u.ac.jp/\~{}watanabe/onlinematerial/f6a.eps and 
  f6b.eps}
\end{figure*}

\paragraph{Space-time slit diagrams}
We have constructed what we call ``multi-slit space-time diagrams''
for all 17 \EBs\ to facilitate quantitative measurement of their
apparent footpoint proper motion, the extension and retraction speeds
of their jet-like protrusions, and the sign and magnitude of their
Doppler signals.  These measurements are non-trivial due to the motion
that the \EBs\ display along the network while flaring up in a
succession of jet-like or thread-like protrusions.
Figure~\ref{fig:slit-image} shows such a diagram for EB-1.  Movie
versions of these diagrams are available in the online material 
(http://www.kwasan.kyoto-u.ac.jp/\~{}watanabe/onlinematerial/).

The measurement procedure is as follows.  Firstly, the approximate jet
direction (N, NE, E, \dots) is determined by eye.  Secondly, a
baseline spatial ``slit'' (called slit~I) of 1\,px width is defined
manually such that it contains the bottom edge, or footpoint, of the
protrusion and tracks its subsequent locations; this track is
generally not straight.  Thirdly, six more slits~II-VII are positioned
by spacing them parallel to slit~I, sharing the same bends, at every
128\,km in the approximate jet direction.  Slit~VII is then the
furthest from the footpoint, at 768\,km from slit~I.  Together, these
slits represent samplings of the jet along its apparent height that
track its footpoint path.  They enable the construction of space-time
plots for the wing brightness $I_{\rm sum}$ and wing Dopplershift
$I_{\rm diff}$.  In the latter, we found it necessary to mask off all
areas that have wing brightness below 1400 (in the arbitrary data
units also used in Figure~\ref{fig:ha-profiles}) because otherwise,
overlying chromospheric fibrils dominate the Doppler signature with
their independent Dopplershifts.  In addition, for these dark features
the Dopplergram definition switches sign, causing extra confusion, as
does their varying viewing angle with respect to the line of sight.
By masking the dark fibrils off, the space-time Doppler signature is
much better constrained to display \EB\ properties only.  Finally,
using these multi-slit space-time diagrams we derived the proper
motion speed of all \EB\ feet, the extension and retraction speeds of
all jets, and the spatial occurrence distributions along all \EBs\ of
their Doppler signatures, as described in the next paragraphs.  The
results are given in Section~\ref{sec:results} and
Table~\ref{tbl:EBs}.

\begin{figure*}[htbp]
\centerline{\includegraphics[width=0.95\textwidth]{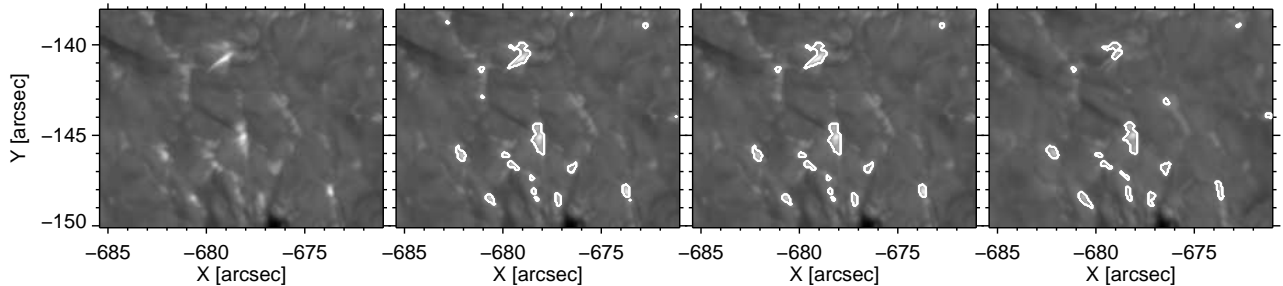}}
  \caption[]{\label{fig:bp-detection}%
    Automatic detection of network bright points using the wing 
    brightness images. {\em First panel\/}: Region~A at
    08:08~UT.  
    The $(X,Y)$ coordinates correspond to Figure~\ref{fig:full-fov}.
    {\em Second panel\/}: Areas over $3\sigma$ brighter than the
    average values are outlined with white contours.  {\em Third
      panel\/}: result after removal of small bright points.  {\em
      Fourth panel\/}: a similar map one minute later, illustrating
    good continuity.}
\end{figure*}

\subsection{Network bright points} \label{sec:BPs}

In order to compare the characteristics between \EBs\ and normal
network bright points (henceforth NBP), we also analyzed the latter in
our data sequence.  Like the \EBs, the NBPs are not distributed
uniformly within the field of view but occur preferentially to the
East of the spot (Figure~\ref{fig:full-fov}).  In order to compare NBP
properties between different types of region
six subregions named A--F were chosen that are identified in the lower
panels of Figure~\ref{fig:full-fov}.  Regions~A, B and C sample the
moat flow around the spot on its North, East, and South
sides. Region~C contains the well-developed part of the penumbra,
whereas Region~D samples the penumbra-free West side of the spot.
Region~E samples the same-polarity plage to the North of the spot,
whereas Region~F samples the opposite-polarity plage to the East of
the spot.

We applied a brightness threshold algorithm to identify NBPs for
statistical analysis using the wing brightness images.  A similar
algorithm was developed for the detection of umbral dots in
\citet{2009ApJ...702.1048W}. 
Note that \EBs\ are also NBPs when categorized per brightness
threshold, but the \EBs\ defined and selected above possess eruptive
jets as well.  The procedure consists of four steps
illustrated in Figure~\ref{fig:bp-detection}:

\begin{enumerate}

\item \emph{Identify NBPs in each image.}  NBPs are defined as areas
  whose wing-sum brightness ($I_{\rm sum}$) is over 3$\sigma$ brighter
  than the average of the pertinent subregion, with $\sigma$ the
  standard deviation. For plage regions a threshold of 5$\sigma$ is
  used instead.  The latter selects only a few NBPs in plage that
  usually lie at the plage periphery.  A lower threshold would select
  much of the plage as NBP, without individual properties or
  measurable morphology.  We then defined the position of each NBP as
  the peak location within the thresholded area.

\item \emph{Remove too small NBPs.}  Too small NBPs covering
  less than 5 pixels (0.18~arcsec in radius) are removed.  Note that
  the diffraction limit is 0.17~arcsec.

\item \emph{Track NBPs.}  A NBP at a subsequent time step is
  considered to mark continued manifestation of a prior NBP if they
  have spatial overlap.  We accepted the cases in which such
  continuation skipped one or two time steps.  When there is no
  overlap within three successive time steps, the NBP is considered to
  have disappeared.

\item \emph{Remove the merging/splitting events.}  The procedures
  described above include some merging or splitting events.  We remove
  these events by selecting the longest-duration event and separate it
  from merging/splitting pieces.  Thus, one of two merging or
  splitting pieces is kept for tracking while the other 
  ends or starts at this time step.
\end{enumerate}

For example in Region~A, a total of 183 NBPs were detected in this
manner.  Seven of these were also selected as \EBs\ in
Section~\ref{sec:EBs}.

\begin{table*}[htb]
\begin{center}
\caption[]{\label{tbl:EBs} \EB\ properties.}
\begin{tabular}{c|c|c|cc|c|cc|c}
\tableline \tableline
EB & jet & lifetime & footpoint & jet & footpoint & extension 
   & retraction & blue/red  \\ 
& direction & [s] & extent\tablenotemark{1} 
   & extent\tablenotemark{1,2}  
   & speed  & speed\tablenotemark{2} 
   & speed\tablenotemark{2} 
   & signal at  \\ 
& &  & [km] & [km] & [\kms] & [\kms] & [\kms] & J:jet, F:footpoint \\ 
\tableline 
1 & SE & 1004 & 820 & 1180 & 1.23 & 5.5 -- 9.8 & 5.9 & J:blue, F:red\\
2 & E & 508 & 360 & 770 & 1.15 & 6.2 & 9.8 & no\\
3 & E & 1302 & 390 & 690 & 1.13 & 6.2 & 5.6 & J:blue\\
4 & E & 713 & 460 & 770 & 1.00 & 7.6 & \dots & J:blue\\
5 & E & 260 & 280 & 1230 & no & 5.7 & 4.6 -- 6.4 & J:blue, F:red\\
6 & E & 266 & 280 & 1120 & no & 11.0 & 8.7 & J:blue, F:red\\
7 & E & 359 & 490 & 870 & 1.60 & 6.4 -- 8.3 & 6.9 &  J:red, F:blue\\ 
8 & E & 682 & 310 & 750 & 1.31 & 3.8 -- 4.2 &10.4 & F: blue, then red\\
9 & E & 322 & 620 & 540 & 1.11 & 23.7 & \dots & J:blue\\
10 & E & 651 & 510 & 850 & 1.44 & 7.2 -- 8.3 &11.9 & J:blue, F:blue\\
11 & E & 428 & 690 & 670 & 0.85 & 18.5 & \dots & J:red\\
12 & NE & 459 & 330 & 1050 & 0.29 & \dots & 8.5 -- 15.1 & F,J: red, then blue\\
13 & N & 279 & 900 & 720 & no & 6.4 -- 11.9 & 6.4 & J:blue, F:red\\
14 & E & 868 & 440 & 950 & 0.27 & 2.8 -- 3.5 & 9.2 & J:red,  F:blue\\
15 & E & 589 & 410 & 800 & no & 2.9 & \dots & no\\ 
16 & SE & 614 & 410 & 930 & no & 6.4 -- 11.9 & 11.9 & J: red, then blue\\
17 & E & 322 & 310 & 640 & no & 6.2 & \dots & J:red, F:blue\\ 
\tableline \tableline
Ave. & E & 566 & 471 & 855 & 0.67 & 8.0 & 8.7 & \ldots\tablenotemark{3} \\
\tableline
\end{tabular}
\begin{minipage}{.7\hsize}

  $^{1}$ The apparent footpoint measured on
  best-seeing images
 
  $^{2}$ If a jet is vertical, which they are roughly, the actual
  lengths and velocities are a factor 1.5 larger
  
  $^{3}$ Too much spread to give a meaningful average

\end{minipage} 
\end{center}
\end{table*}

\section{Results} \label{sec:results}

\subsection{Ellerman bomb properties}

Our measurements of the 17 selected \EBs\ are given in
Table~\ref{tbl:EBs}.  Their footpoint travel lengths along the network and   
their jet extents were measured on best-seeing images.  The resulting
jet extents and the jet extension and retraction speeds are apparent
lengths and velocities.  If the jets are vertical, which they roughly
seem to be, the real values are larger by a factor $1/\mu=1.5$.  The
mean apparent jet extent is close to one Mm, so the jets stick out
well above the photosphere.  The mean lifetime is about 10 minutes.

\paragraph{Footpoint motion}
The wing brightness distribution across slit~II in the space-time
diagrams such as the one for EB-1 in Figure~\ref{fig:slit-image} was
used for the measurement of footpoint proper motion because this slit
usually contains the brightest part of the bright point at the foot of
an \EB.  When there is proper motion, the space-time diagram shows an
inclined bright lane, as is the case in Figure~\ref{fig:slit-image}.
The inclination was then measured and converted into apparent speed.
The direction of motion is always radially away from the spot. 
The average speed of footpoint motion of \EBs\ is 
0.67\,\kms. 

\paragraph{Extension and retraction speed}
We identified apparent jet extensions or retractions using the wing
brightness panels of the multi-slit space-time diagrams together with
the corresponding movie.  Some examples are marked in
Figure~\ref{fig:slit-image} with white dotted lines.  Their slope
gives the extension/retraction speed by dividing the pertinent slit
separation in units of their 128\,km spacing by the travel time.  The
typical value is 5\,--\,10\,\kms, without systematic difference
between the extension and retraction speeds.  Of course, the actual
gas speed depends on the projection; for vertical jets the measured
apparent speeds should be multiplied by $1/\mu=1.5$.

\paragraph{Dopplershift}
We found interesting Dopplershift signatures in the wing 
Dopplergram space-time charts in 15 cases out of 17 \EBs.  
EB-1 in Figure~\ref{fig:slit-image} starts with subtle blueshift and later
displays simultaneous redshift at the footpoint NBP and blueshift near
the jet top during 500\,s -- 750\,s.  Three more cases show such
combination of footpoint redshift with jet blueshift.  The pattern
suggests the presence of a bi-directional jet, with downflow at its
bottom and upflow at its top.  The blueshifts tend to last longer than
the redshifts.  In three cases we found the opposite pattern: downflow
at the top and upflow at the bottom.  Sometimes an EB alternates
between the two patterns during its lifetime.  In the two cases where
the jet is clearly accompanied by a surge (EB-8 and EB-12, discussed
below), the latter's Dopplershift obscures the underlying \EB\ motion.
In two cases (EB-2 and EB-15) we could not find \EB\ 
associated Dopplershifts, probably due to interweaving upflow 
and downflow components or seeing perturbation.     

\begin{figure}[htbp]
  \centerline{%
    \includegraphics[width=0.45\columnwidth]{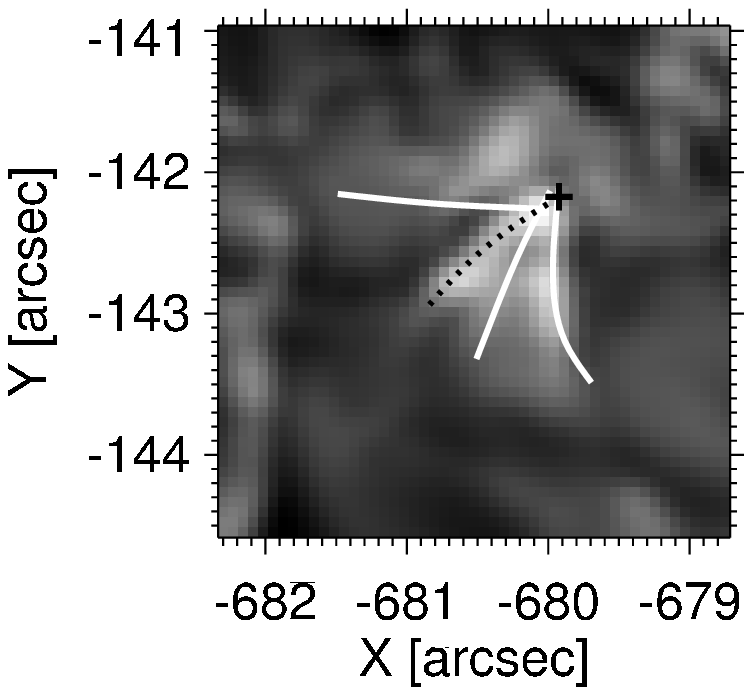}  
    \includegraphics[width=0.45\columnwidth]{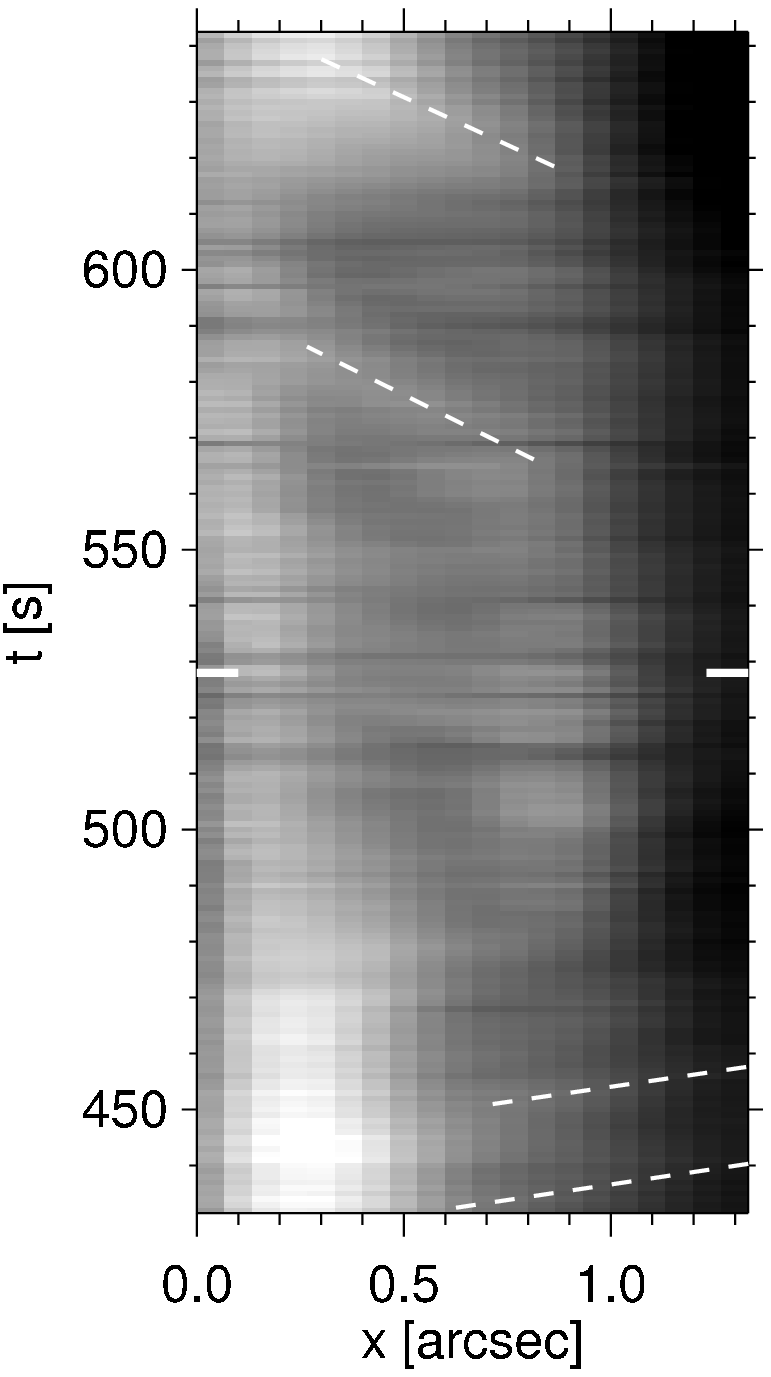}}
  \caption[]{\label{fig:wb-measurement} {\it Left\/}: measurement
    tracks through \EB\ EB-1 in the wide-band data at
    08:04\,UT. The $(X,Y)$ coordinates correspond to
    Figure~\ref{fig:full-fov}.
    {\it Right\/}: time-space diagram along
    the black dotted track in the lefthand image, with the elapsed time from
    the first appearance plotted vertically.  
    The 0~arcsec location is indicated in the right panel with a black plus sign.
    The slanted dotted lines specify apparent-speed estimations, while the 
    white tick marks at t$\approx$530\,s 
    indicate the timestep at which sample data is shown in the right
    panel.}
\end{figure}

\begin{figure}[htbp]
  \centerline{\includegraphics[width=0.8\columnwidth]{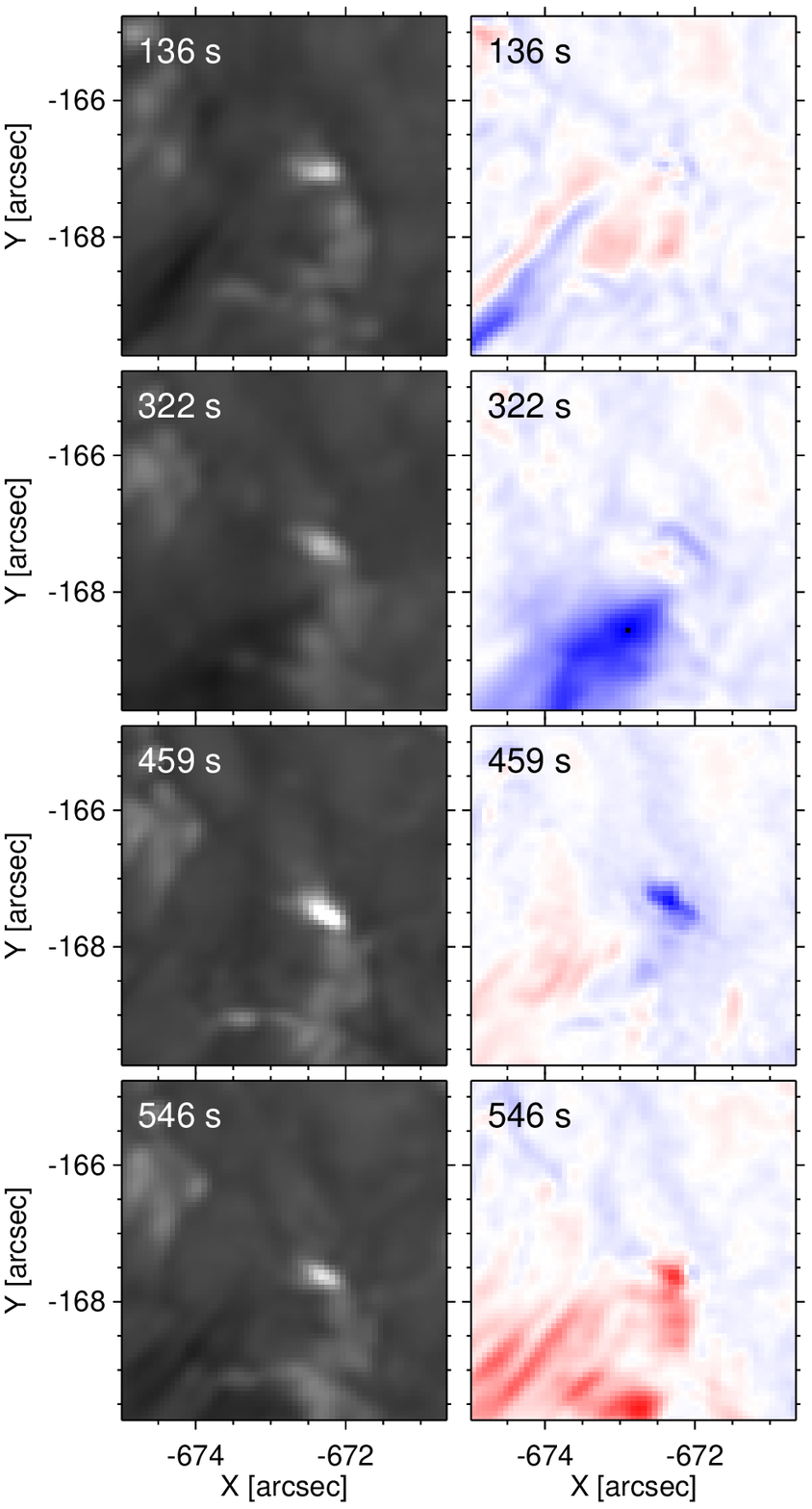}}
  \caption[]{\label{fig:EB-8-surge} Temporal evolution of Ellerman
    bomb EB-8 in wing-brightness ({\it left\/}) and wing-Dopplergram images
    ({\it right\/}).  The
    numbers in each panel specify the elapsed time from its first
    appearance.  The $(X,Y)$ coordinates correspond to
    Figure~\ref{fig:full-fov}.  The limb is to the left.  This
    \EB\ seems to initiate a dark extension to its South-East
    (lower-left direction), with alternating Dopplershift signature (where
    blue color implies downflow for the absorption feature.)}
\end{figure}

\paragraph{Fast variations}
Figure~\ref{fig:cut-fast} and the wide-band movie in the online material
demonstrate that, at the 1-s cadence of the wide-band image sequence,
very fast variations are observed.  
As residual high-frequency seeing effects might contaminate the sample of
real fast variations, a low-pass filter with a cut-off of 50\,\kms\ 
was applied to the wide-band data. 
In addition, only those variations that were visible in more than one 
subsequent frame were considered for the analysis described below, while 
residual seeing effects are typically only visible for one frame and
furthermore random in nature.

A more detailed analysis of the
fast variations was performed for EB-1.  First, the cutout subfield
frames were cross-correlated to remove EB-1's proper motion.
Time-space diagrams were then extracted along paths that track in time 
either the jet top or single bright features inside the \EB.  The
lefthand panel of Figure~\ref{fig:wb-measurement} shows the selected
paths overlaid on a good quality EB-1 image.  The righthand panel
shows part of the time-space diagram along the dashed path.  The
timeslices enable measurement of the apparent proper motion speed for
particular features that move along these paths, as well as the jet
extension and retraction speed.  Four such measurements have been
overlaid as dashed lines in the right panel in
Figure~\ref{fig:wb-measurement} to illustrate the procedure.  The
upper two concern downdraft features, both with an apparent speed
around 18\,\kms, whereas the lower two are updrafts, with apparent
speeds close to 59\,\kms.

In this manner eleven displacement speed measurements were obtained
for EB-1.  The typical timescale for jet-like features to
extend/retract is 25--40\,s, but isolated brightness features can
exist on timescales of 2--7\,s.  Six out of eleven cases display
updrafts and the rest display downdrafts.  The apparent updraft speeds
range between 11--60\,\kms\ with the majority at the higher values,
while the apparent downdraft speed range between 7--18\,\kms.

\paragraph{Surges}
Figure~\ref{fig:EB-8-surge} shows the case of EB-8 which displays
special temporal evolution.  First, the \EB\ appears and seems to
initiate surge activation on its South-East side.  This surge appears
repetitively during the observations, with alternating negative and
positive $I_{\rm diff}$.  Note that, since the surge is observed as an
absorption feature in \Halpha, the Dopplershift sign is reversed with
respect to the \EB\ which is observed in emission, so that for the
surge, positive wing difference $I_{\rm diff}$ implies upflow along
the line of sight, i.e., blueshift although it is color-coded red in
Figure~\ref{fig:EB-8-surge}.  The surge exhibits strong positive
$I_{\rm diff}$ corresponding to an upflow of $\approx 40$\,\kms
(measured from the \Halpha\ line-center Doppler shift) at around
322\,s after the appearance of EB-8.  About 1~min after the surge
disappears, negative $I_{\rm diff}$ (downflow) appeared at the
position of the \EB, and 1~min after that positive $I_{\rm diff}$
(upflow) replaced the negative $I_{\rm diff}$.  Simultaneously, the
surge location was also dominated by redshift.

EB-12 is similarly adjacent to an \Halpha\ surge on its South side, 
which also shows alternating positive and negative $I_{\rm diff}$. 

\begin{figure}[htbp] 
  \centerline{\includegraphics[width=0.75\columnwidth]{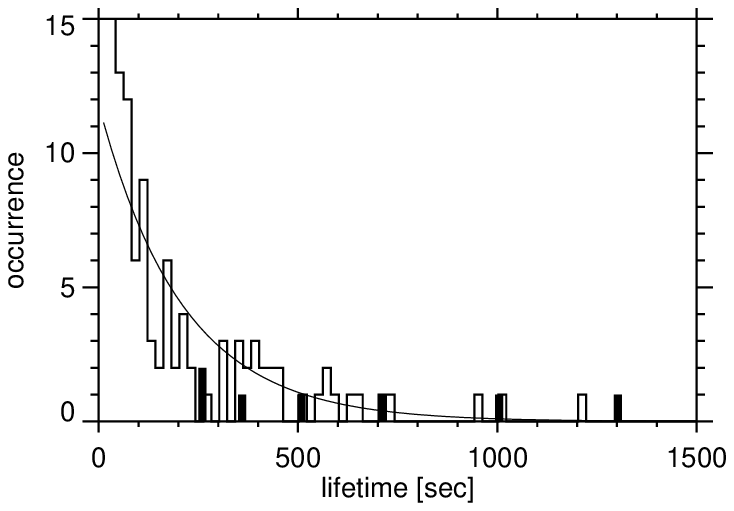}}
  \centerline{\includegraphics[width=0.75\columnwidth]{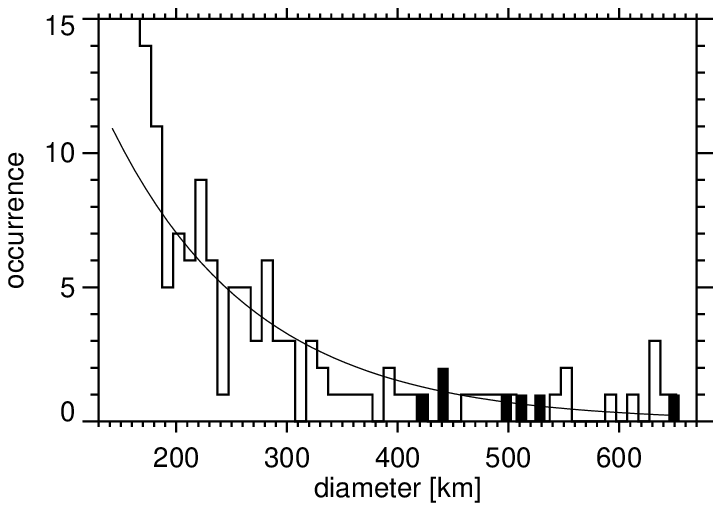}}
  \centerline{\includegraphics[width=0.75\columnwidth]{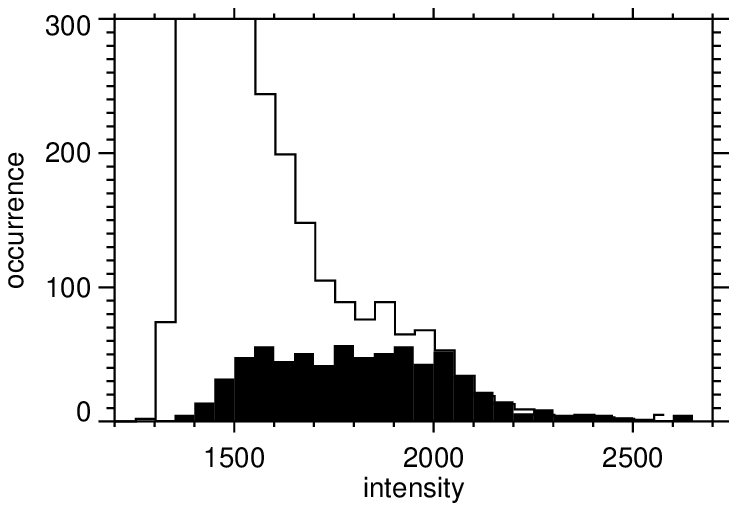}}
  \caption[]{\label{fig:histograms}%
    Histograms of bright-point lifetime, diameter, and brightness in Region~A.
    The 7 \EBs\ found in this region are entered as black
    bins. The curves in the upper two panels are fits with exponential
    decay functions $y={\rm const.} \times {\rm exp}(-x/C)$. }
\end{figure}

\subsection{\EBs\ versus normal NBPs}

Figure~\ref{fig:histograms} shows histograms of the lifetime,
diameter, and wing brightness of NBPs in Region~A.  The seven NBPs
that are also \EBs\ are entered as black bins. The diameter is
calculated by assuming circular shape, converting its area into the
corresponding circle diameter, and taking the average over the NBP
lifetime.  The histogram for brightness contains values from every
time step, making the number of samples large.  The detection
thresholds for the lifetime and diameter measurement were 6.2\,s and
135\,km, respectively.  The brightness threshold is set by the NBP
detection algorithm (Section~\ref{sec:BPs}).

As for other automated detection methods
(\cite{1997A&A...328..682S}; 
\cite{2009ApJ...702.1048W}), 
these histograms show exponential decrease.  We fit the first two
histograms with functions $y \propto {\rm exp}(-x/C)$ with $C$ the
characteristic value.  In the case of Region~A, a characteristic
diameter of 273\,km and characteristic lifetime of 222\,s were
obtained.  The only striking difference between \EBs\ and normal NBPs
is in the brightness panel, \EBs\ being extraordinary bright in the
\Halpha\ wings---commensurate with their definition---but with overlap
during their startup and decay phases which makes the presence of a
jet a necessary discriminator between \EBs\ and the brightest NBPs.
We conclude that, apart from their brightness, \EBs\ are just like
non-eruptive NBPs.

Table~\ref{tbl:NBPs} shows the average proper motion speed and
occurrence rate of the NBPs in the six subregions.  The average proper
motion speed was determined from NBPs with a lifetime longer than
60\,s by dividing the distance between the disappearance and
appearance locations by its lifetime, and taking the average of those
for each region.  The occurrence rate is defined by the average NBP
occurrence frequency within one square arcsec area per minute.

\begin{table*}[htbp]
\begin{center}
\caption{\label{tbl:NBPs} NBP properties}
\begin{tabular}{c|c|c|c|c|c}
 \tableline \tableline
Region & Total Number& Characteristic & Characteristic & Average proper & Occurence rate \\ 
           & of NBPs detected & lifetime [s] & diameter [km] & speed [{\kms}] & [arcsec$^{-2}$ min$^{-1}$]  \\ \tableline
A 	& $183$	& $222$ 		& $273$ 		& $0.88$ 	& $0.028$ \\ 
B 	& $141$ 	& $232$ 		& $214$ 		& $0.75$ 	& $0.030$  \\ 
C 	& $206$	& $240$ 		& $255$ 		& $0.83$ 	& $0.033$  \\ 
\hline 
D 	& $89$ 	& $173$ 		& $225$ 		& $0.59$ 	& $0.039$ \\
\hline 
E 	& $41$ 	& $111$	         & $205$ 	        & $\ldots^{*}$ 	& $0.002$ \\ 
F 	& $6$ 	& $\ldots^{*}$ 	& $\ldots^{*}$ 	& $\ldots^{*}$	& $0.001$ \\ \tableline
\end{tabular}\\
${}^{*}$Too few samples to calculate 
\end{center}
\end{table*}

Table~\ref{tbl:NBPs} suggests a subdivision into three types of
region.  Firstly, Regions A, B and C can be characterized as moat flow
regions in which NBPs are relatively long-lived and large, moving with
the same speed as that of the moat flow.  
Second, he occurrence rate in Region~D, covering the outside of a 
penumbra-free area, is highest, but the NBPs do not move much and 
their lifetimes are relatively short. 
The appearance of NBPs in this region follows the
intergranular lane morphology.  Thirdly, Regions E and F sampling
plage contain only a few individual NBPs that fit our detection
algorithm.  These lie mostly at the plage peripheries.  The other
bright features within the plage are too crowded for our detection
scheme so that we cannot measure their characteristic lifetimes and
diameters, but they tend to be smaller and more static.

\section{Discussion}\label{sec:discussion}

\paragraph{\EB\ occurrence, type, mechanism}
Most \EBs\ in our data appear on the East side of the spot that has a
well-developed penumbra.  They seem to take part in the moat flow that
emanates from the penumbra in their fast proper motion along the
network (see
\cite{2007ApJ...660L.165V, 2008ApJ...679..900V} 
for evidence that moat flows require penumbrae).  In particular, the
repetitive brightening of jet threads and the jet direction 
(away from disk center in the limbward field of view) suggests that new flux with
near-vertical orientation arrives at the network as driven by the moat
flow, and reconnects with the pre-existing, also largely vertical,
network fields.  The brightenings start low and extend rapidly upwards
in jet-like flames, sometimes evolving into bi-directional jets.  
The flows at the location of \EBs\ change sign alternately 
and the jet lengths extend and retract repetitively.  This
behavior indicates reconnection that takes place again and again 
by encountering pre-existing magnetic field.  The highly
supersonic apparent feature speeds measured from the wideband data
suggest phase speed rather than material motion, reminiscent of the
shear motion described by
\citet{1998ApJ...509..435V}. 

The recent literature discusses multiple types of \EBs.  In particular,
Pariat \etal, \citet{2004ApJ...614.1099P}, 
\citet{2007A&A...473..279P}, and \citet{2008ApJ...684..736W} 
found evidence for undulating ``serpentine'' field
topologies with \EBs\ formed at the dip of field lines 
(called ``bald patches'' in the French literature), where the fields
are nearly antiparallel.  The fact that \EBs\ are reported to only
occur in emerging flux regions speaks for an antiparallel reconnection
scenario.  The repetitivity of the bright-thread appearance while the
\EB\ migrates along the network may indeed result from sea-serpent
field topology, but located in the photosphere rather than extending
to the chromosphere.

Since we have no simultaneous magnetogram data at similar angular
resolution, we can only speculate about the magnetic topology of our
\EBs.  There may be opposite-polarity field emergence with direct or
with separatrix/separator reconnection (cf.\,
\cite{2004A&A...428..595P}), 
or there may be component reconnection between unipolar fields that
shear along each other in differential moat flows.  In our case, we
did observe considerable field topology changes at the East side of
the spot during the two magnetogram sequences taken before and after
the \Halpha\ sequence, but this gives only an indication that further
field emergence may well have taken place during the \Halpha\ sequence.
We think it most likely that the \EBs\ in our data require a field
topology that produces near-vertical reconnection deep in the
photosphere, and we suspect that the moat flow is a key operator.  The
strong differential moat flow combined with the presence of nearby
bipolar plage might suffice as condition to produce our type of \EB.

Recently,
\citet{2010ApJ...724.1083G} 
reported an SST plus {\em Hinode\/} observation of a flux emergence
region exhibiting an event much like our \EBs\ and accompanied by an
\Halpha\ surge.  They attribute it, with the help of simulation
examples, to reconnection due to interaction between emerging flux and
the pre-existing chromospheric and coronal field.  In contrast, in our
\EBs\ the reconnective interaction seems to be fully defined within
the photosphere, although two of our seventeen \EBs\ do have an
accompanying surge.

\paragraph{Photospheric formation}
All \EBs\ in our \Halpha\ images originate from intergranular lanes,
i.e., from regular granulation in the deep photosphere that is sampled
by the \Halpha\ wings apart from occasional Doppler-shifted absorption
features (\cite{2006A&A...449.1209L}; 
\cite{2009ApJ...705..272R}, 
Paper~I).  The reversed granulation at heights $h = 100-200$~km is
seen in \Halpha\ only closer to line center in extremely quiet
fibril-free regions.
All our \EBs\ lie below the chromospheric fibril canopy.  \Halpha\ is
commonly thought to be chromospheric; for example,
\citet{2002ApJ...575..506G} 
described their \EBs\ as low-chromosphere features on the basis of
their detection in filtergrams at $\Delta \lambda \!=\! -0.8$\,\AA\ from
\Halpha\ line center
which suggests that these were actually as photospheric as ours.
Therefore, our observations strengthen the conclusion of
\citet{2002ApJ...575..506G} 
that \EBs\ are probably a signature of low-altitude reconnection.

\paragraph{Similarity to chromospheric anemone jets}
All our \EBs\ have extended jets; in fact, the jet presence became a
selection criterion after our initial movie inspections.
\citet{2007Sci...318.1591S} 
reported the existence of so-called ``chromospheric anemone jets''
seen in \CaIIH\ with the {\it Hinode} Solar Optical Telescope.  They
have lengths of 2000--5000\,km and apparent extension speeds of
10--20\,\kms.  They are found preferentially in mixed-polarity regions
(\cite{2010PASJ...62..901M}). 
Particularly in a sunspot moat,
\citet{2010PASJ...62..901M} 
found nine such chromospheric anemone jets with lifetimes ranging
between 30--320\,s, in good agreement with our \EB\ durations.  Some
of our \EBs\ also show the inversed Y-shape which is characteristic
for anemone jets, and also the apparent jet extension speeds are
comparable.
\citet{2010PASJ...62..901M} 
attributed the jet extension speed to the Alfv\'en velocity in the low
chromosphere; we note that the Alfv\'en velocity in the photosphere 
(at field strength 1000~Gauss and density 10$^{17}$\, cm$^{-3}$) is  
comparable.  The major difference is that the \EB\ jets tend to reach
smaller heights.  
Nevertheless, we suggest that our type of \EBs\ and
chromospheric anemone jets represent the same phenomenon, with the
difference in \Halpha\ and \CaIIH\ extent perhaps due to the
difference in fibril visibility in the two diagnostics.

\paragraph{\EB\ heating}
Our data suggest strongly that our \EBs\ are caused by 
magnetic reconnection of strong near-vertical network fields with also
near-vertical field that is probably moat-flow driven into the
network.  Undoubtedly there is sizable heating.  There are attempts to
model the bright \Halpha\ wings of \EBs\ by adding appropriate
temperature humps to standard models in one-dimensional modeling or
inversion
(\cite{2006SoPh..235...75S}; 
\cite{2010MmSAI..81..646B}), 
but these are likely to underestimate the upward extent of the heating
by trying to also reproduce the dark \Halpha\ core which in reality is
caused by opaque chromospheric fibrils that are not taken into
account.  We suspect that exceedingly large thermal line width, as
proposed by \citet{2008PASJ...60...95M} 
and particular to the Balmer lines due to the small atomic mass of
hydrogen, may be an important aspect.  \Halpha\ line synthesis in
numerical MHD simulations including encounter or close-encounter
reconnection is obviously desirable.

\section{Conclusion}  \label{sec:conclusion}
The imaging spectroscopy presented in this paper is far superior to
any \Halpha\ observation of \EBs\ in the literature.  It yielded
unprecedented spatial and temporal resolution that we exploited to
describe the structure and evolution of 17 \EBs\ seen in the \Halpha\
wings.  They are all rooted deeply in the magnetic network on the side
of a sunspot with strong moat flow, and they all have upright jets
that reach considerable height, although these jets remain shielded by
overlying chromospheric fibrils at \Halpha\ line center.

Although multiple types of \EB\ and corresponding field configurations
are reported in the literature, the ones reported here seem all to be
photospheric events that are most likely explained by successive
reconnection of strong vertical network fields with incoming, also
vertical, field that is driven into the network by the moat flow.

The \Halpha\ observations presented in this paper have one obvious
shortcoming: there was no simultaneous polarimetry performed that
would provide corresponding magnetograms at similar angular
resolution.  Such data is required to study the topology and evolution
of the magnetic fields at and around \EB\ sites to ascertain what
happens magnetically, in particular to establish the presence and
action of field emergence and cancelation and to chart the field
polarity.  Note that such mapping should be photospheric.
Chromospheric magnetometry is currently a grail of solar
spectropolarimetry, but for \EBs\ as the ones discussed here it would
diagnose the field of overlying fibrils, not the bombs or their
triggering, and be irrelevant unless the \EB\ sends off a measurable
surge.

During 2010, the Oslo group obtained new SST/CRISP sequences that
combine \Halpha\ and Ca~8542 line scans with \FeI~6302
spectropolarimetry of a sunspot in an emerging flux region yet closer
to the limb.  These data are well suited for further study, presently
underway, of the \EB\ phenomenon.

\acknowledgments We thank our colleagues at the Kwasan and Hida
Observatories and at Kyoto University and Oslo University for fruitful
discussions.  This research was supported by a Grant-in-Aid for JSPS
Fellows, a Grant-in-Aid for the Global COE Program ``The Next
Generation of Physics, Spun from Universality and Emergence'' of the
Ministry of Education, Culture, Sports, Science and Technology (MEXT)
of Japan and by the Research Council of Norway through grants
177336/V30 and 191814/V30.  The Swedish 1-m Solar Telescope is
operated on the island of La Palma by the Institute for Solar Physics
of the Royal Swedish Academy of Sciences in the Spanish Observatorio
del Roque de los Muchachos of the Instituto de Astrof{\'\i}sica de
Canarias.  R.J.~Rutten visited Japan as a guest of NAOJ at Mitaka and
acknowledges hospitality at Kyoto University and Hida Observatory.


\end{document}